\documentclass[prl, aps, twocolumn, floatfix, superscriptaddress, amsmath,  longbibliography]{revtex4-1}
\usepackage{graphicx}
\usepackage{graphics}
\usepackage{mathtools}
\usepackage{float}
\usepackage{amssymb}
\usepackage{bbold}

\usepackage[mathlines]{lineno}

\usepackage[dvipsnames]{xcolor}
\definecolor{darkblue}{rgb}{0.0, 0.0, 0.75}

\usepackage{color}
\usepackage[colorlinks=true,
            linkcolor=darkblue,
            urlcolor=darkblue,
            citecolor=darkblue]{hyperref}

	\usepackage[normalem]{ulem} 
	\definecolor{mgreen}{RGB}{1,123,0}

\usepackage{siunitx}

\def \br{{\bf r}}

\def \mref{{\mathrm{ref} }}

\def \mdc{\mathrm{dc}}

\def \mHz{\mathrm{Hz}}

\def \mms{\mathrm{ms}}

\begin{document}
%

\title{Shapiro spikes in driven quantum fluids}
\author{Vijay Pal Singh}
\thanks{Corresponding author: vijay.singh@tii.ae}
\affiliation{Quantum Research Centre, Technology Innovation Institute, Abu Dhabi, UAE}
\author{Ludwig Mathey}
\affiliation{Zentrum f\"ur Optische Quantentechnologien and Institut f\"ur  Quantenphysik, Universit\"at Hamburg, 22761 Hamburg, Germany}
\affiliation{The Hamburg Centre for Ultrafast Imaging, Luruper Chaussee 149, Hamburg 22761, Germany}
\author{Luigi Amico}
\affiliation{Quantum Research Centre, Technology Innovation Institute, Abu Dhabi, UAE}
\affiliation{Dipartimento di Fisica e Astronomia, Universit\`a di Catania, Via S. Sofia 64, 95123 Catania, Italy}
\affiliation{INFN-Sezione di Catania, Via S. Sofia 64, 95127 Catania, Italy}
\date{\today}
%

\begin{abstract}
Periodic driving of a Josephson junction produces resonant transport signatures known as Shapiro spikes, conventionally described by effective circuit models. 
Here, we demonstrate Shapiro spikes in a neutral quantum fluid using simulations of a potential-biased atomic Josephson junction. 
The dc-current resonances follow the expected Bessel-function scaling, including the suppression and revival of individual resonances with increasing modulation strength. 
We uncover their microscopic mechanism by showing that the phase-winding frequency develops finite plateaus at integer multiples of the drive frequency. 
These phase-locked states are sustained by quantized-vortex nucleation and the associated $2\pi$ phase slips. 
Remarkably, within a locking plateau, the dc current and density transfer can reverse direction 
while the phase-winding frequency remains locked to the drive, giving rise to negative-mobility transport. 
Our results establish a microscopic link between vortex-mediated phase slips, Shapiro resonances, and anomalous transport in a neutral quantum fluid, revealing how quantized vortices convert resonant transport into phase-locked dynamics with negative mobility. 
\end{abstract}
\maketitle
%
%

 The Josephson effect across a weak link between two coherent reservoirs is a paradigmatic manifestation of macroscopic
quantum coherence \cite{Josephson1962,GrossNotes,barone1983josephson,Likharev1979}. 
When the junction is periodically driven at frequency $\omega$, its nonlinear phase dynamics 
can generate pronounced resonant transport responses. 
In the conventional current-biased setting, these appear as Shapiro steps in the voltage, 
reflecting synchronization of the phase dynamics to integer multiples of the drive frequency 
\cite{Shapiro1963,Grimes1968}. 
A complementary regime emerges when the conjugate potential difference, 
rather than the current, is externally imposed. 
In this potential-biased configuration, resonances occur when the Josephson frequency $\omega_0$ set by the dc bias becomes commensurate with the drive frequency, $\omega_0= n \omega$, where $n$ is an integer. 
These resonances manifest as sharp structures in the dc current, 
commonly referred to as \emph{Shapiro spikes} \cite{barone1983josephson,Galda2015}. 
Despite their long-standing theoretical prediction, the microscopic dynamics underlying this regime---and, in particular, how resonance, dissipation and directed transport combine to produce the current response---remain largely unexplored.

Atomic Josephson junctions offer a particularly natural route to this regime. 
Weak links in ultracold bosonic and fermionic superfluids have enabled studies of Josephson oscillations, critical transport, 
current–phase relations and dissipative phase-slip dynamics \cite{Cataliotti2001,Albiez2005, Levy2007, LeBlanc2011, Eckel2014_2, Valtolina2015, Spagnolli2017,
Pigneur2018,Kwon2020,Luick2020,Pace2021,SinghAJJ2025,Gan2026}, 
while recent experiments have extended this correspondence to conventional Shapiro steps \cite{DelPace2025,Bernhart2025}, 
including extensive theoretical work on dc and ac driven dynamics 
\cite{Smerzi1997,Raghavan1999,Giovanazzi2000, Meier2001,Polo2018,Zaccanti2019, Singh2020jj, Piazza2010,Piazza2011, Xhani2020, Xhani2020NJP, Gabriel2023, Jahrling2025, SinghShapiro,  Singh2026JPA, SinghPANS,
Geier2026, Singh_hyst}.
Crucially, neutral-atom systems allow the relative energies of spatially separated reservoirs to be prescribed directly using shaped optical potentials \cite{Hueck2018,Luick2020,Grani2025,Diego2026}. The dc and ac components of the reservoir-energy difference can therefore be controlled independently while the weak-link barrier remains stationary, providing direct access to a potential-biased geometry that is difficult to isolate in electronic junctions because of the influence of leads and the surrounding circuit environment. 
At the same time, the spatially resolved density, phase and vortex dynamics of the quantum fluid provide microscopic observables 
that are normally inaccessible in solid-state devices, 
allowing the transport response to be connected directly to the underlying microscopic many-body dynamics 
\cite{Husmann2015,Valtolina2015,Burchianti2018,DelPace2025}.
Driven two-mode models were considered in Refs. \cite{Eckardt2005, Grond2011}.

Here we demonstrate Shapiro spikes in a potential-biased atomic quantum fluid using semiclassical-field simulations. 
We impose controlled dc and ac energy offsets between two reservoirs while keeping the weak link stationary. 
Periodic modulation generates a hierarchy of sharp, dispersive dc-current resonances at commensurate biases $\omega_0=n\omega$.  
Their amplitudes follow the characteristic Bessel-function hierarchy of a driven Josephson mode, 
including non-monotonic suppression and revival with modulation strength and a collapse across different drive frequencies and resonance orders.

   Microscopic access to the quantum fluid reveals that Shapiro spikes are accompanied by 
resonantly enhanced vortex nucleation. 
The resulting quantized $2\pi$ phase slips provide a dissipative channel for releasing the accumulated phase difference and sustain phase locking, manifested as finite plateaus in the phase-winding frequency at integer multiples of the drive frequency. This vortex-mediated mechanism is absent from standard circuit descriptions of Shapiro spikes. 
Within a locking plateau, the phase-winding frequency remains fixed even as the bias changes the intra-cycle phase dynamics and redistributes the relative contributions of forward and backward current. 
This redistribution can reverse the time-averaged dc current and density transfer, 
producing a negative-mobility branch within the locked state.
Our results connect the macroscopic Shapiro-spike response to quantized vortex-mediated phase slips, 
phase locking, and anomalous superfluid transport.

\begin{figure}[t!]
\includegraphics[width=0.98\linewidth]{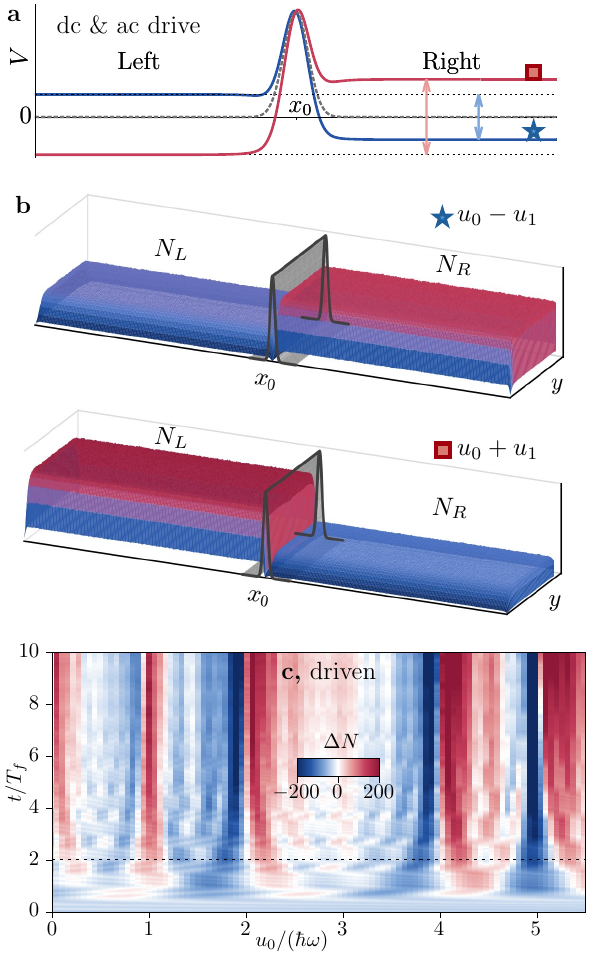}
\caption{\textbf{System, protocol and resulting many-body resonant dynamics.}  
\textbf{a,} Sketch of the potential landscape. 
A localized Gaussian barrier centered at $x_0$ forms an atomic Josephson junction between left and right reservoirs. 
Their relative energies are modulated according to $u(t)=u_0-u_1\cos(\omega t)$, 
where $u_0$ is the dc bias, $u_1$ is the ac bias amplitude, and $\omega = 2\pi f$ is the drive frequency.
The reservoir-energy difference therefore oscillates between $u_0+u_1$ (red) and $u_0-u_1$ (blue), while the tunnel barrier remains stationary. 
\textbf{b,} Corresponding density distributions at the two extrema of the combined dc-ac bias. 
$N_L$ ($N_R$) denotes the atom number of the left (right) reservoir.
The density is suppressed in the junction region by the barrier (grey), 
whose height $V_0$ and width $w$ control the weak-link coupling. 
\textbf{c,} Driven imbalance dynamics, $\Delta N(t)=N_L(t)-N_R(t)$, as a function of $u_0/\mu$ 
for $V_0/\mu=9$, $f=60\,{\rm Hz}$ and $u_1/\mu=0.9$, 
where $\mu$ is the mean-field energy and $T_f= 1/f$ is the drive period. 
The initial transient for $t/T_f < 2$ (horizontal dashed line) is excluded from the analysis of the driven response.
} 
\label{Fig:system}
\end{figure}

\begin{figure*}[t]
\includegraphics[width=1.0\linewidth]{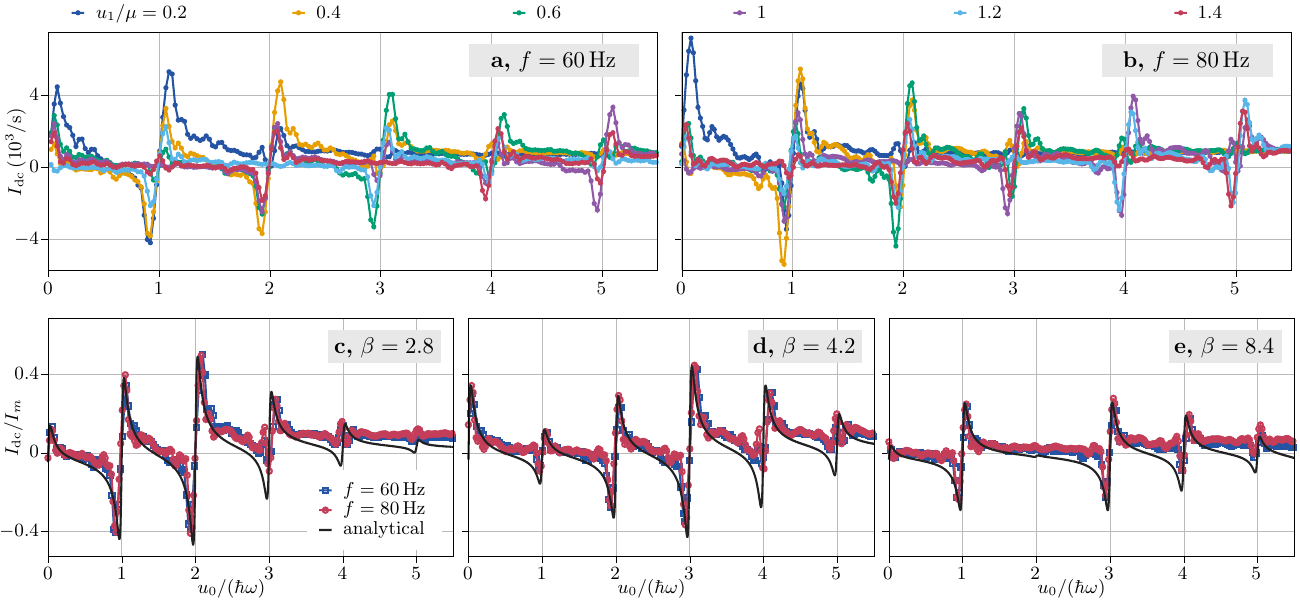}
\caption{\textbf{Shapiro spikes in a potential-biased atomic Josephson junction.}  
\textbf{a-b,} Current--bias characteristics $I_{\rm dc}$--$u_0$ under periodic driving for increasing modulation amplitudes $u_1/\mu$ at drive frequencies $f=60\,{\rm Hz}$ and 
$80\,{\rm Hz}$, respectively. 
Sharp dispersive current resonances emerge near the commensurability conditions $u_0=n\hbar\omega$. 
\textbf{c-e,} Comparison between the atomic-junction response and the collective Josephson model (equation \ref{eq:ac2}) for scaled modulation strengths $\beta=u_1/(\hbar \omega)=2.8$, $4.2$ and $8.4$, respectively. 
Currents are normalized by $I_m$, the maximum dc current obtained under a static potential bias (Methods).
The model uses the relaxation rate $\Gamma/(2\pi) \simeq 2\,{\rm Hz}$ extracted from the driven response at $\beta=2.1$ (Methods), and captures the resonance positions, antisymmetric shapes and their non-monotonic evolution with $\beta$. 
} 
\label{Fig:spikes}
\end{figure*}

\section{System and protocol}
We consider a homogeneous two-dimensional Bose gas confined to a box of dimensions $L_x\times L_y$.  
A localized Gaussian barrier divides the system into left and right reservoirs, forming an atomic Josephson junction (AJJ) (Fig.~\ref{Fig:system}a).
We simulate the dynamics using a classical-field approach within the truncated Wigner approximation \cite{Blakie2008, Polkovnikov2010, Singh2016, Singh2020sound}, starting from the Hamiltonian
\begin{equation} \label{eq_hamil}
\hat{H}_{0} = \int \mathrm{d}{\bf r} \Big[ \frac{\hbar^2}{2m} \nabla \hat{\psi}^\dagger({\bf r}) \cdot \nabla \hat{\psi}({\bf r})  + \frac{g}{2} \hat{\psi}^\dagger({\bf r})\hat{\psi}^\dagger({\bf r})\hat{\psi}({\bf r})\hat{\psi}({\bf r})\Big].
\end{equation}
$\hat{\psi}$ ($\hat{\psi}^\dagger$) is the bosonic annihilation (creation) operator. 
The interaction $g=\tilde{g} \hbar^2/m$ is given in terms of the dimensionless parameter 
$\tilde{g} =  \sqrt{8 \pi} a_s/\ell_z$, 
where $m$ is the mass, $a_s$ is the $s$-wave scattering length, and 
$\ell_z$ is the transverse harmonic-oscillator length. 
We sample the initial many-body fields $\psi (\br,t=0)$ from a grand-canonical ensemble 
at chemical potential $\mu$ and temperature $T$ using a classical Metropolis algorithm, 
thereby incorporating equilibrium fluctuations beyond mean-field description (Methods).
Each initial state is then propagated under the classical equations of motion
\begin{align}\label{eq:eom}
 i \hbar \dot{\psi}(\br, t) = \Bigl(  - \frac{\hbar^2}{2m} \nabla^2 + V(\br, t) + g|\psi|^2 \Bigr) \psi(\br, t). 
\end{align}
The weak link and reservoir bias are generated by the external potential 
\begin{equation}\label{eq:potential}
V(\br, t) = V_0 (t)\exp\!\left[-\frac{2(x-x_0)^2}{w^2}\right]  + \frac{u(t)}{2} \tanh\!\left(\frac{x-x_0}{w}\right),
\end{equation}
where $V_0(t)$, $w$, and $x_0 = L_x/2$ denote the barrier height, width, and position, respectively. 
The second term directly controls the potential-energy difference between the two reservoirs.
For concreteness, we consider a gas of $^{6}{\rm Li}_2$ molecules with density $n\simeq 5\,\mu{\rm m}^{-2}$, 
interaction parameter $\tilde g=0.1$, condensate fraction of approximately $90\%$, 
and box dimensions $L_x\times L_y=512\times26.5\,\mu{\rm m}^2$ (Methods). 
We define the dimensionless barrier height by $\tilde V_0\equiv V_0/\mu$, 
where $\mu=gn$ is the mean-field energy. 
The barrier width is kept fixed at $w/\xi=1.05$, 
where $\xi=\hbar/\sqrt{2mgn}$ is the healing length.

\begin{figure*}[t]
\includegraphics[width=1.0\linewidth]{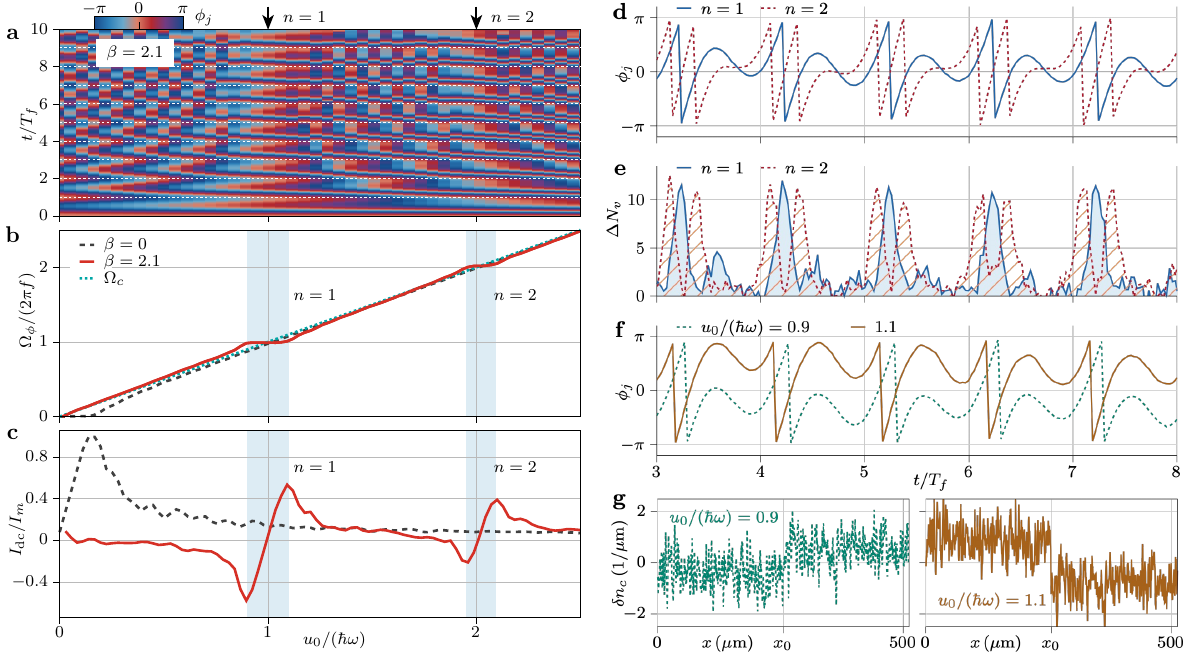}
\caption{\textbf{Phase locking, vortex-mediated phase slips and rectified transport.}
\textbf{a,} Time evolution of the wrapped junction phase difference $\phi_j(t)$ under ac driving.
Away from resonance the phase exhibits irregular running-phase dynamics, whereas near resonance 
it becomes phase locked with the external drive (arrows). 
\textbf{b,} Dimensionless phase-winding frequency $n_{\rm eff}=\Omega_\phi/(2\pi f)$ without ($\beta=0$) and with ($\beta=2.1$) ac driving. Shaded regions mark the $n=1$ and $n=2$ phase-locking plateaus.
The dotted line indicates the time-averaged phase-winding rate predicted by the effective circuit model, $\Omega_c = \omega_0$, 
which follows the dc-bias-induced Josephson frequency without developing locking plateaus.
\textbf{c,} Corresponding dc current--bias characteristic $I_{\rm dc}$--$u_0$, showing resonant Shapiro spikes at the phase-locking conditions.
\textbf{d,} Representative locked phase dynamics at the first ($n=1$, solid) and second ($n=2$, dashed) resonances, corresponding to one and two $2\pi$ phase windings per drive period, respectively.
\textbf{e,} Simultaneous vortex response $\Delta N_v= N_v(t) - N_\mref$ for the $n=1$ and $n=2$ phase-locked states,  
where $N_\mref$ is the reference vortex number in the absence of dc and ac bias. 
\textbf{f,} Junction-phase dynamics near the lower- and upper-bias edges of the first locking plateau, $u_0/(\hbar\omega)=0.9$ (dashed) and $1.1$ (solid), respectively. 
Both trajectories remain locked at $n_{\rm eff}=1$, but their intra-cycle phase distributions yield opposite signs of the time-averaged Josephson current.
\textbf{g,}  Time-averaged density change $\delta n_c = \langle \int n(x, y,t) dy - n_\mref(x) \rangle_t$ for the two trajectories in \textbf{f}, showing opposite density redistribution across the junction. 
The reference line density $n_\mref(x)$ is obtained in the absence of dc-ac bias, and $x_0$ denotes the barrier position.
} 
\label{Fig:phase}
\end{figure*}

Starting from an unbiased system, $u(t)=0$, we slowly ramp the barrier to its final value, 
producing two weakly coupled reservoirs with atom numbers $N_L$ and $N_R$ (Fig.~\ref{Fig:system}a). 
We then impose a combined dc and ac reservoir-energy difference 
\begin{equation} \label{eq:bias} 
u(t)=u_0-u_1\cos(\omega t), 
\end{equation} 
where $u_0$ is the dc bias, $u_1$ is the modulation amplitude, and $\omega=2\pi f$ is the drive frequency. 
Crucially, this protocol modulates the relative reservoir energies while keeping the tunnel barrier fixed 
(Fig.~\ref{Fig:system}a,b). It therefore creates genuinely potential-biased Josephson junction, 
distinct from self-induced oscillations \cite{Carraro2026} or moving-barrier protocols in which transport is generated mechanically \cite{SinghShapiro}. 
We characterize the transport by the atom imbalance $\Delta N(t)=N_L(t)-N_R(t)$ across the junction. 
Under static dc bias, the imbalance response develops a pronounced maximum at finite $u_0$ (Methods and Fig.~\ref{Fig:phase}c). 
Periodic modulation qualitatively restructures the dynamics: 
a sequence of resonant structures emerges at biases commensurate with the drive, $u_0 \simeq n \hbar \omega$, with $n=0,1,2,\ldots$, as visible directly in the time-dependent imbalance (Fig.~\ref{Fig:system}c).  
These resonances occur when the Josephson frequency set by the dc bias becomes commensurate with the external modulation and constitute the potential-biased counterpart of conventional Shapiro steps. Here, however, the potential difference is imposed externally and the conjugate atom current is detected, so the commensurability appears as a resonant dc-current response—the Shapiro-spike regime explored below. 
As we show, in the quantum fluid this resonant response is accompanied by genuine phase locking: the phase-winding frequency develops finite plateaus at integer multiples of the drive frequency.

\section{Shapiro spikes}
To determine the current-potential ($I_\mdc$--$u_0$) characteristic of the junction, 
we define the dc current as $I_{\rm dc}=\langle I(t)\rangle_t$, with $I(t)=d(\Delta N)/dt$, 
where the average is taken over the interval between $2$ and $10$ cycles (Fig.~\ref{Fig:system}c).  
Figure~\ref{Fig:spikes}a,b shows $I_\mdc$ as a function of the dc bias $u_0$ for increasing modulation amplitude $u_1$ and two drive frequencies $f$. 
Periodic driving produces sharp current resonances near $u_0\simeq n\hbar\omega$. 
Rather than simple peaks, the resonances exhibit a dispersive, approximately antisymmetric shape, with positive and negative extrema of comparable magnitude. 
Their amplitude varies non-monotonically with the modulation strength and drive frequency. 
The responses obtained at different drive frequencies collapse when expressed in terms of the scaled modulation strength $\beta= u_1/(\hbar \omega)$ (Fig.~\ref{Fig:spikes}c-e). 
This scaling reflects the universal Bessel-function dependence $J_n(\beta)$ of the Shapiro resonances, 
as shown below. 

 The resonances obtained here reflect the response of a realistic quantum fluid and 
 therefore depart from the idealized delta-function Shapiro spikes of textbook Josephson models. 
 Dissipation broadens the response into finite-width, dispersive antisymmetric resonances, 
 with vortex-mediated phase relaxation providing the microscopic dissipative channel in our system.
The broadening increases as the barrier height is reduced, 
consistent with enhanced vortex nucleation (Methods). 
Higher barriers therefore produce sharper spikes and improve spectral resolution, particularly at low drive frequencies. 
At the same time, the critical current decreases with increasing barrier height, reducing the absolute magnitude of the resonant signal. 
An intermediate barrier height therefore provides an optimal balance 
between spectral sharpness and experimentally detectable current amplitude.

This trade-off is not determined by barrier height alone: the barrier width also controls the tunneling amplitude, and the two parameters can be tuned jointly to tailor the junction coupling \cite{Singh2020jj}. 
The relative influence of barrier height and width depends on the barrier geometries 
accessible in a given atomic platform. 
Indeed, sharp spikes are also obtained for a barrier height of $\tilde{V}_0=2$ at widths $w/\xi=4$ and $5$  (Methods and Fig. \ref{Fig:spikesVs2}), demonstrating the robustness of the resonance response to changes in barrier geometry. Such tunability allows the trade-off between spectral resolution and signal strength to be optimized for different atomic platforms \cite{SinghAJJ2025, Diego2026}.

The $I_\mdc$--$u_0$ characteristic additionally contains a weak approximately linear background, visible as an overall tilt at larger bias (Fig.~\ref{Fig:spikes}c-e). 
This contribution reflects the dissipative, approximately Ohmic component of the transport, 
whereas the pronounced resonant structure is dominated by the Josephson contribution (Methods). 
The resulting response thus consists of a weak dissipative background superimposed with sharply enhanced, phase-coherent dc transport at the resonance  conditions.

 To understand the origin of this response, we treat the Josephson mode as a collective degree of freedom governed by the Josephson relations, while incorporating dissipation through a phenomenological relaxation rate $\Gamma$. 
We define the relaxation-weighted dc current as $I_\mdc = \Gamma \int_0^\infty e^{- \Gamma t} I_s(t) dt$, with $I_s(t) = I_c \sin \phi (t)$. 
Despite its minimal form, this description captures the main features of the microscopic transport response (Methods). 
For combined dc-ac bias, the Josephson relation gives 
$\phi(t)=\phi_0+\omega_0 t - \beta\sin(\omega t)$, where $\phi_0$ is the initial phase difference and $\omega_0=u_0/\hbar$. 
Expanding the phase modulation into harmonics of the drive yields  
\begin{equation}\label{eq:ac}
\frac{I_\mdc}{I_c}  = \sum_{n=-\infty}^{\infty} J_n(\beta)  \frac{\Gamma^2 \sin \phi_0 + \Gamma (\omega_0- n \omega) \cos \phi_0}{\Gamma^2 + (\omega_0- n \omega)^2}. 
\end{equation}
The response therefore contains both symmetric Lorentzian and antisymmetric dispersive contributions, 
whose relative weights are controlled by the initial phase $\phi_0$. 
For $\phi_0=\pi/2$, the response is purely Lorentzian, with a characteristic width set by $\Gamma$. 
As the damping decreases, the resonance becomes increasingly narrow and localized around $\omega_0= n\omega$, approaching the singular, infinitely sharp Shapiro response of the idealized system. 
For the initial condition used here, $\phi_0=0$, the Lorentzian contribution vanishes and the response becomes purely dispersive,
\begin{equation}\label{eq:ac2}
\frac{I_\mdc}{I_c}  = \sum_{n} J_n(\beta)  \frac{\Gamma (\omega_0- n \omega)}{\Gamma^2 + (\omega_0- n \omega)^2}. 
\end{equation}
This expression directly explains the characteristic resonance shape obtained in the atomic junction. 
Each harmonic is centered at the Shapiro condition $\omega_0=n\omega$ and changes sign across the resonance center. Its extrema occur at 
$\omega_0 = n\omega \pm \Gamma$, with magnitude $I_c |J_n(\beta)|/2$. 
The Bessel-function structure naturally accounts for the non-monotonic evolution of the resonances with increasing drive strength, including their suppression near the zeros of $J_n(\beta)$ and subsequent re-emergence.

We note that $\Gamma$ plays a qualitatively different role for Shapiro spikes than for Shapiro steps: while for Shapiro steps, damping effects lead to smooth-out of the plateaus, for Shapiro spikes $\Gamma$ is responsible for the resonant behavior itself, as also discussed below. 
Since $\Gamma$ varies with the barrier height, 
we determine its value  by fitting the driven response to equation \ref{eq:ac2} (Methods). 
Using the extracted $\Gamma$, the collective model reproduces the integer resonance positions, antisymmetric shapes, characteristic widths and non-monotonic dependence on $\beta$ observed in the simulations (Fig.~\ref{Fig:spikes}c--e).
Higher-order resonances become progressively more prominent with increasing modulation strength, in agreement with the redistribution of spectral weight among the Bessel harmonics.
The agreement establishes the Shapiro origin of the current spikes and shows that their macroscopic structure is captured by a dissipative collective Josephson mode, while their microscopic realization in the quantum fluid is governed by the phase and vortex dynamics discussed below.

\begin{figure}[t]
\includegraphics[width=1.0\linewidth]{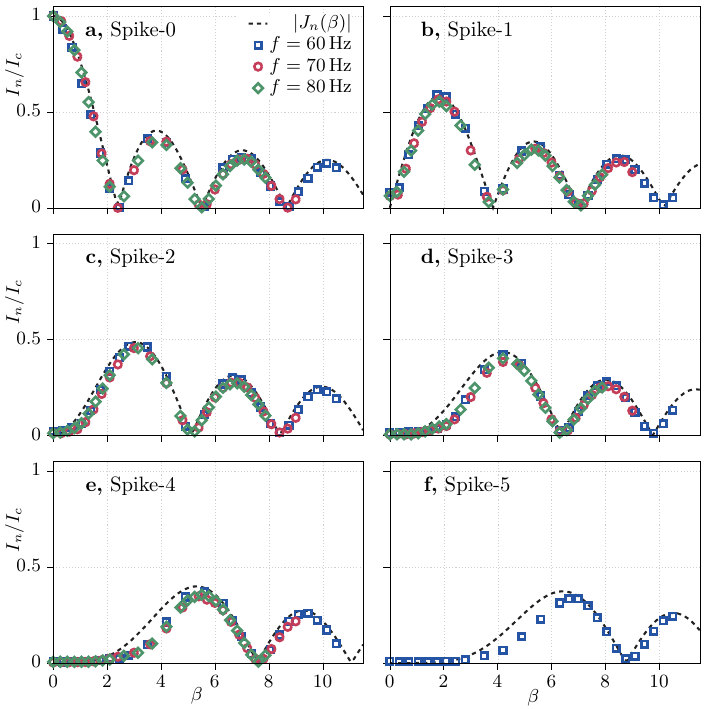}
\caption{\textbf{Bessel-function scaling of the Shapiro spikes.}
Normalized amplitudes $I_n/I_c$ of the $n=0$--$5$ Shapiro resonances as a function of 
the scaled modulation strength $\beta=u_1/(\hbar\omega)$ for drive frequencies $f=60$, $70$, and $80~{\rm Hz}$ at fixed barrier height $\tilde{V}_0=7$. 
Symbols show the atomic-junction simulations, while dashed curves show the analytical prediction $|J_n(\beta)|$.
Data obtained at different drive frequencies collapse onto the corresponding Bessel-function curves, 
capturing successive suppression and revival of the resonance.
}
\label{Fig:bessel}
\end{figure}
The analytical model (equation \ref{eq:ac2}) predicts that the amplitude of the $n$-th Shapiro resonance is governed by the Bessel function $J_n(\beta)$, with $\beta=u_1/(\hbar \omega)$. 
To test this prediction, we extract the resonance amplitudes $I_n$ from the $I_{\rm dc}$--$u_0$ characteristics over a range of modulation strengths and drive frequencies, 
defining $I_n$ as the full peak-to-peak current difference between the two extrema of each dispersive resonance.  
Figure~\ref{Fig:bessel} shows $I_n/I_c$ as a function of $\beta$ for resonance orders $n=0,\ldots,5$. 
Data obtained at different drive frequencies collapse onto common curves and reproduce the successive maxima, minima, and zeros of $|J_n(\beta)|$, 
demonstrating that the Shapiro response is controlled by the scaled modulation strength rather than by $u_1$ and $\omega$ independently.

Together with the resonance condition $u_0 \simeq n \hbar \omega$, this collapse shows that both the location and amplitudes of the Shapiro spikes are governed by the dimensionless variables $u_0/(\hbar\omega)$ and $\beta$ identified by the collective Josephson description. 
The agreement is strongest for the lower-order resonances, 
while systematic deviations emerge at higher orders, where spatial excitations and many-body dynamics neglected by the effective model become increasingly relevant. 
Nevertheless, the persistence of the Bessel hierarchy across multiple resonance orders and drive frequencies provides a stringent quantitative signature of Shapiro resonant behavior in the microscopic atomic junction.

\section{Phase dynamics}
To characterize the resonant dynamics, we analyze the junction phase difference $\phi_j(t)=\phi(x_0-1.5w,t)-\phi(x_0+1.5w,t)$,
evaluated on the two sides of the barrier from the transversely averaged complex field (Fig.~\ref{Fig:phase}a).  
Away from resonance, the ac-driven junction remains in a running-phase regime in which the phase winds 
continuously and, when represented within the interval $[-\pi, \pi]$, repeatedly crosses the branch cut. 
The resulting dynamics are generally irregular. 
Near resonance, by contrast, the phase evolution reorganizes into a regular pattern synchronized with the external drive, and this locking persists over a finite interval of dc bias.

We quantify this behavior through the normalized phase-winding frequency $n_{\rm eff} = \Omega_\phi/\omega$, 
where $\Omega_\phi$ is obtained from the slope of a linear fit to the unwrapped junction phase
(Fig.~\ref{Fig:phase}b).
Without periodic driving, $n_{\rm eff}$ varies continuously with dc bias in the running-phase regime, with $\hbar \Omega_\phi \simeq u_0 $, demonstrating the ac Josephson relation in the potential-biased geometry. 
The dc bias therefore directly sets the phase-winding frequency of the junction. 

Periodic driving qualitatively changes this response. 
Instead of varying continuously with $u_0$, the phase-winding frequency develops finite plateaus at integer values $n_{\rm eff}=n$. 
On the $n$-th plateau, the junction accumulates approximately $2\pi n$ of phase per drive period, corresponding to one winding per period for $n = 1$, two for $n = 2$, and so forth (Fig.~\ref{Fig:phase}d).
The first two winding-frequency plateaus coincide with the corresponding Shapiro resonances  (Fig.~\ref{Fig:phase}c), demonstrating that, in the quantum liquid, the resonant dc-current response is accompanied by genuine frequency locking of the junction phase dynamics. 
The Shapiro spikes therefore provide the transport signature of a phase-locked state.

The microscopic origin of these winding events is revealed by the vortex dynamics (Methods).
Figure~\ref{Fig:phase}e shows pronounced peaks in vortex nucleation at the first and second current resonances, 
coincident with the $n = 1$ and $n = 2$ synchronized states. 
Their correlation with the quantized phase winding identifies the resonant $2\pi$ relaxation events as vortex-mediated phase slips. 
Remarkably, if vortices are neglected as it happens in effective circuit-model descriptions, 
the finite-locking plateaus are not recovered (Fig.~\ref{Fig:phase}b).
Quantized vortex nucleation therefore provides a microscopic dissipative channel through which 
the accumulated phase difference is periodically released, 
allowing the junction to remain synchronized with the external drive.
This intrinsic dissipation stabilizes phase locking over a finite bias interval, 
replacing the singular resonance of an ideal nondissipative junction by a finite-width Shapiro response. In a quantum fluid, such finite locking intervals also make the resonances directly resolvable in transport measurements.

The phase slips themselves, however, do not directly generate the dc Josephson current, 
since a $2\pi$ phase change leaves the instantaneous Josephson current unchanged. 
The finite dc component instead originates from the asymmetric phase evolution 
between successive slips. 
In particular, the junction can spend unequal fractions of each drive cycle in regions where $\sin \phi_j$ is positive and negative, producing a nonzero time average of $\sin \phi_j$. 
Periodic driving therefore rectifies the oscillatory phase dynamics into a directed dc superflow.
This mechanism is particularly clear for two phase-locked trajectories near the lower- 
and upper-bias edges of the first synchronization plateau  (Fig.~\ref{Fig:phase}f).  
Both remain locked at $n_{\rm eff}=1$ and accumulate one $2\pi$ phase winding per drive period, 
yet their intra-cycle phase distributions are shifted toward opposite sides of $\phi_j=0$. 
Near the lower-bias edge, the phase spends a larger fraction of each cycle in regions with $\sin \phi_j < 0$, yielding a negative time-averaged Josephson current. 
Near the upper-bias edge, it preferentially samples regions with $\sin \phi_j > 0$, 
yielding a positive average current.  
The phase-winding frequency therefore remains locked while the rectified Josephson current reverses across the synchronization plateau. 

Remarkably, this reversal extends to the total dc transport. 
Near the lower-bias edge, $I_\mdc <0$ despite the applied dc bias remaining positive.
The junction therefore supports net atom transport opposite to the imposed static bias, 
demonstrating a negative-mobility regime of the type anticipated in driven nonlinear systems  \cite{Reichhardt2017}. 
The sign reversal is directly encoded in the dispersive Shapiro resonance: 
its two lobes correspond to synchronized states with the same quantized winding number but oppositely directed dc currents. The Shapiro spike thus provides a direct transport signature of negative mobility in a quantum fluid.

The corresponding real-space signature is shown in Fig.~\ref{Fig:phase}g. 
States near the two edges of the synchronization plateau exhibit opposite time-averaged density redistributions across the junction. 
The reservoir supplying the net current is depleted, 
while density accumulates in the receiving reservoir. 
Near the lower-bias edge, the negative dc current corresponds to net right-to-left transport, 
whereas near the upper-bias edge the positive current corresponds to left-to-right transport. 
The reversal of the density redistribution therefore provides an independent, spatially resolved signature of the current reversal and, on the negative-current branch, of transport opposing the applied dc bias.

\section{Conclusion and outlook}
We have established the Shapiro-spike regime in a potential-biased atomic Josephson junction from its many-body dynamics. 
The combined dc and ac reservoir-energy bias produces dispersive resonances in the dc current at integer commensurabilities between the dc bias and drive frequency.
Remarkably,  these resonances  occur as a many-body synchronization phenomenon in which the phase-winding frequency is pinned to an integer multiple of the drive  frequency.  
Microscopically, this synchronization is sustained by resonantly enhanced nucleation of quantized vortices. 
The resulting $2\pi$ phase slips release the accumulated phase difference and 
provide the dissipation required to maintain synchronization over a finite bias interval. 
Within a locked state, the time-averaged current and  density transfer can reverse direction 
even as the phase-winding frequency remains locked, giving rise to negative mobility. 
Shapiro spikes therefore emerge as vortex-mediated synchronized states whose transport properties
are governed by  microscopic phase-slip dynamics. 

This picture opens several directions for future work. 
The simultaneous access to current, density, phase, and vortices in atomic junctions provides a direct route to experimentally resolving both the Shapiro-spike resonances and their underlying phase-locking dynamics \cite{SinghShapiro, DelPace2025, Bernhart2025}.
Because the resonance condition $u_0=n\hbar\omega$ directly links the applied modulation frequency to the energy bias, 
Shapiro spikes could also enable frequency-to-potential conversion (e.g., potential calibration) and controlled atomtronic transport \cite{Amico2022}. 
More fundamentally, our results establish a setting in which quantized vortex excitations mediate the dissipative phase slips that  transform a resonant transport into a synchronization dynamics. 
Extending this framework to fermionic superfluids and coupled junctions could reveal how collective modes, vortex excitations, and many-body correlations shape synchronized quantum transport.

\section{Acknowledgments}
L.M. acknowledges funding by the Deutsche Forschungsgemeinschaft (DFG) in the excellence cluster  `Advanced Imaging of Matter’ - EXC 2056 - project ID 390715994 and by ERDF of the European
Union and `Fonds of the Hamburg Ministry of Science,
Research, Equalities and Districts (BWFGB)’.
\section{Author contributions}
V.P.S. designed and performed research as well as carried out data analysis.  
L.M. and L.A. contributed to the interpretation of the results.  
V.P.S. wrote the paper with inputs from all coauthors.

\section{Data availability}  
Data supporting the findings of this study are available from the corresponding author upon reasonable request.

\section{Competing interests} 
The authors declare no competing interests.




\section{METHODS}

\textbf{Classical-field simulation details.} 
We consider a uniform bosonic condensate confined in a box potential of dimensions $L_x\times L_y=512\times26.5\,\mu{\rm m}^2$, which we discretize on a $1024\times53$ lattice with spacing $l=0.5\,\mu{\rm m}$. 
The lattice spacing is chosen smaller than both the de Broglie wavelength 
and the healing length, ensuring that the continuum limit is fulfilled \cite{Mora2003}.
We incorporate equilibrium fluctuations by sampling the initial classical fields from the grand-canonical Boltzmann distribution using a Metropolis algorithm. 
At each Monte Carlo update, a lattice site is selected at random and the real and imaginary 
parts of the local field are independently perturbed by normally distributed increments. 
The width of the distribution is set to maintain an acceptance probability of approximately $50\%$. 
We first perform approximately $10^5$ updates to equilibrate the field. 
Independent initial configurations $\psi (\br,t=0)$ are then generated by separating successive samples by more than $5000$ updates per lattice site, suppressing correlations between realizations. 
Each sampled filed is subsequently propagated according to the classical equations of motion 
to obtain the time evolution $\psi (\br,t)$.

\begin{figure*}[t]
\includegraphics[width=1.0\linewidth]{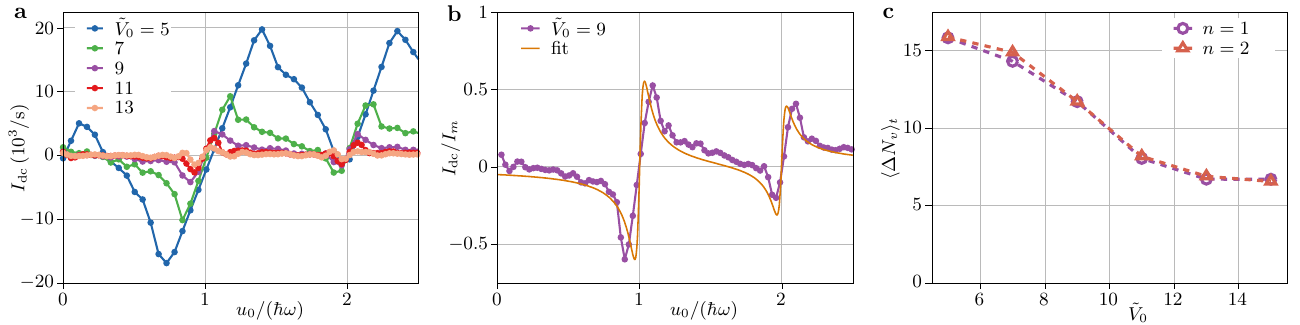}
\caption{\textbf{Characterizing the junction response under an ac bias.}
\textbf{a,} $I_\mdc$--$u_0$ characteristics for an ac driven junction at drive frequency $f=60\, \mHz$ and scaled modulation strength $\beta=2.1$, shown for several barrier heights $\tilde{V}_0$ and width $w/\xi= 1.05$. 
\textbf{b,} $I_\mdc$--$u_0$ characteristic for $\tilde{V}_0=9$, together with a fit to the collective Josephson model (equation \ref{eq:mac2}), yielding an effective damping rate $\Gamma/(2\pi) \approx 2\, \mHz$. 
\textbf{c,} Time-averaged vortex number $ \langle \Delta N_v \rangle_t$ as a function of 
$\tilde{V}_0$ at the first (circles) and second (triangles) Shapiro resonances. 
Here, $\Delta N_v= N_v(t) - N_\mref$, where $N_v(t)$ is the total number of vortices and  $N_\mref=N_v(0)$.   
}
\label{Fig:EDac}
\end{figure*}

\textbf{Vortex detection.} 
We identify vortices from the phase winding of the classical field around each elementary lattice plaquette of size $l\times l$. 
Writing $\theta(x, y)=\mathrm{arg}[\psi(x, y)]$, the phase difference between neighboring sites is wrapped to the principal interval $(-\pi, \pi]$.
For each plaquette, we calculate the circulation $
\delta \theta_\Box = \delta_x \theta(x,y) + \delta_y\theta(x+l,y)+\delta_x\theta(x+l,y+l)+\delta_y\theta(x,y+l)$, 
where $\delta_{x} \theta(x,y)$ and  $\delta_{y} \theta(x,y)$ denote the wrapped phase differences along the $x$ and $y$ directions, respectively. 
$\delta \theta_\Box = 2\pi$ identifies a vortex, whereas $\delta \theta_\Box= -2\pi$ identifies an antivortex. 
The total number at time $t$ is then obtained by summing the absolute phase windings over all plaquettes, $N_v(t) = \sum |\delta \theta_\Box/(2\pi)|$.  
We show the time evolution of $\Delta N_v= N_v(t) - N_\mref$ at first and second Shapiro spikes in Fig. \ref{Fig:phase}e, where $N_\mref= N_v(0)$. During each driving cycle, a single and double vortex nucleation peaks emerge at first and second spikes, respectively. By averaging over the time evolution, we obtain a time-averaged vortex number $\langle  \Delta N_v \rangle_t$ (Fig. \ref{Fig:EDac}c), showing that number of vortices in the nucleation peak decreases with increasing barrier height.

\begin{figure*}[t]
\includegraphics[width=1.0\linewidth]{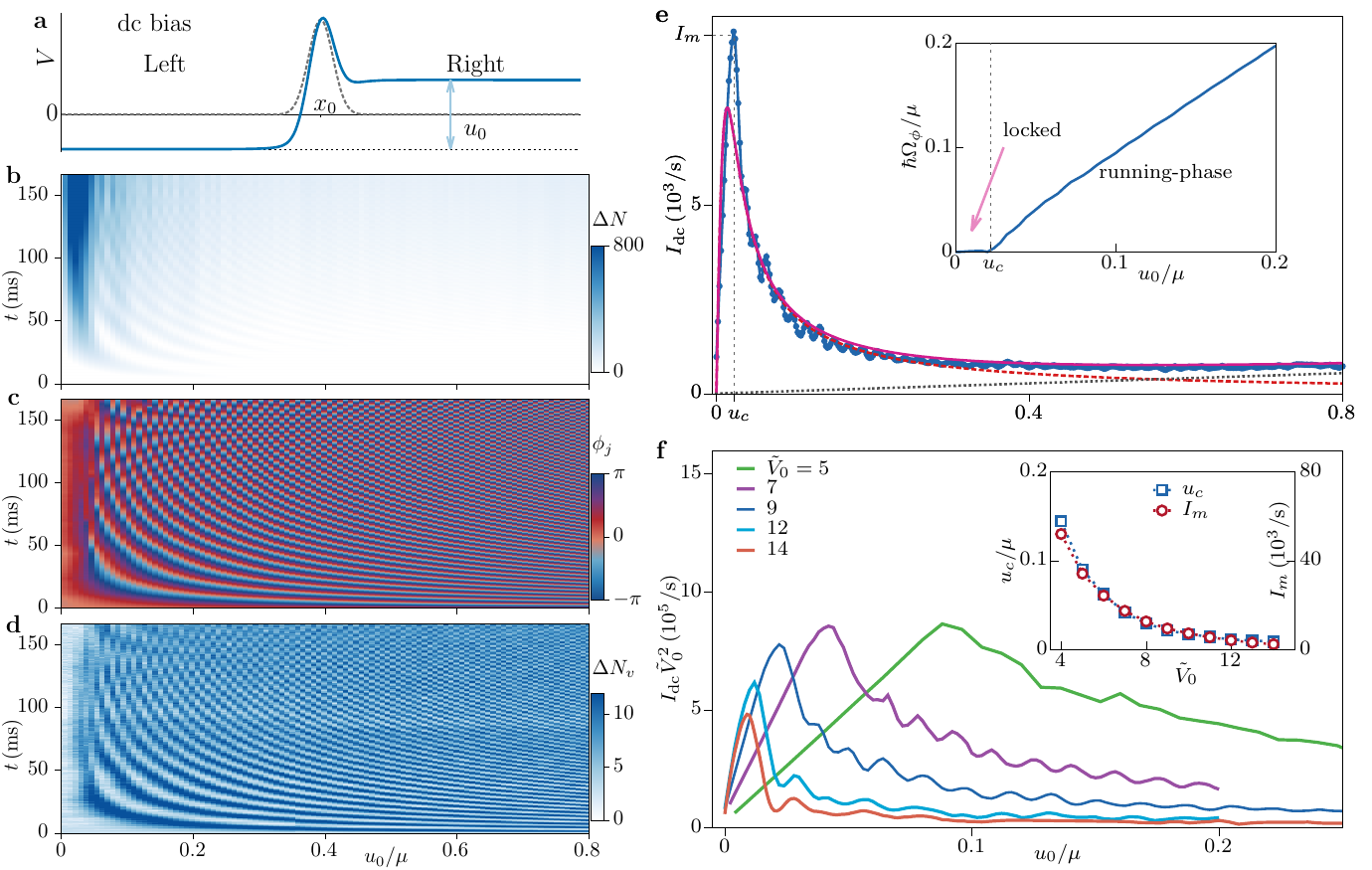}
\caption{\textbf{Characterizing the response of a dc-biased atomic junction.}
\textbf{a,} Sketch of the junction under a dc potential bias $u_0$, 
which imposes an energy difference between the two reservoirs. 
\textbf{b,} Time evolution of the imbalance $\Delta N(t)=N_L(t)-N_R(t)$ as a function of $u_0/\mu$ for barrier height $\tilde V_0\equiv V_0/\mu=9$, where $\mu$ is the mean-field energy. 
\textbf{c-d,} Corresponding wrapped junction phase difference $\phi_j(t)$ and total vortex number $\Delta N_v(t) = N_v(t) - N_\mref$, where  $N_\mref = N_v(0)$. 
\textbf{e,} Resulting dc current--bias characteristic $I_\mdc$--$u_0$.
The current reaches a maximum $I_m$ at the critical bias $u_c$ and decreases as the junction enters the running-phase regime. 
The red dashed line shows the collective-model prediction (equation \ref{eq:adc2}), whereas the red solid line includes an additional weak Ohmic contribution, shown separately by the black dotted line. 
Inset: mean phase-winding frequency $\Omega_\phi$, distinguishing the phase-locked ($u_0<u_c$) and running-phase ($u_0>u_c$) regimes. 
\textbf{f,} $I_{\rm dc}$--$u_0$ characteristics for several barrier heights $\tilde V_0$. 
Inset: extracted critical bias $u_c$ and maximum current $I_m$ as a function of $\tilde V_0$. 
}
\label{Fig:EDdc}
\end{figure*}
\begin{figure}[t]
\includegraphics[width=1.0\linewidth]{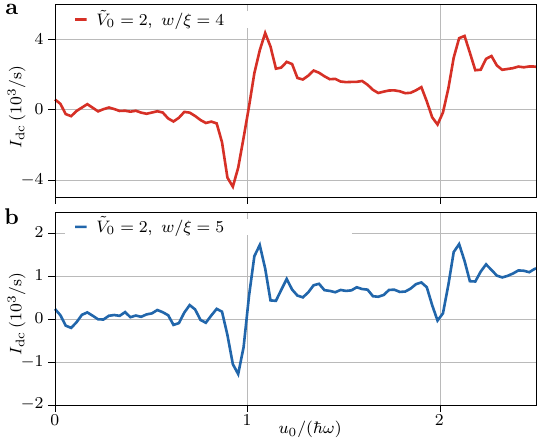}
\caption{\textbf{Dependence on barrier width.}
\textbf{a-b,} $I_{\mathrm{dc}}$ as a function of the dc energy bias $u_0$ for barrier height $\tilde{V}_0=2$ and widths $w/\xi=4$ (\textbf{a}) and $5$ (\textbf{b}). The drive frequency is $f=60~{\rm Hz}$ and $\beta=2.1$. 
}
\label{Fig:spikesVs2}
\end{figure}
\begin{figure*}[t]
\includegraphics[width=1.0\linewidth]{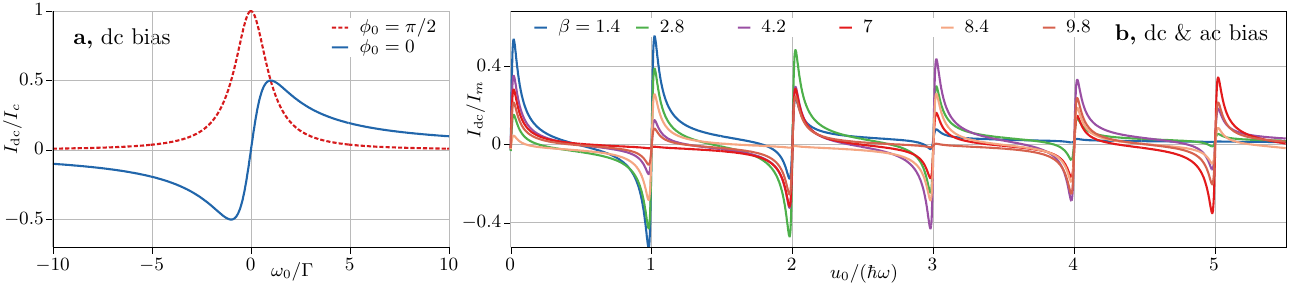}
\caption{\textbf{Collective Josephson-junction model.}
\textbf{a,} Analytical dc-bias response from equation \ref{eq:adc} for initial phase differences $\phi_0=\pi/2$ (dashed) and $\phi_0=0$ (solid). 
The dc current $I_\mdc$ is normalized by the critical current $I_c$, 
and the dc-bias frequency $\omega_0$ is normalized by the relaxation rate $\Gamma$. 
The $\phi_0=\pi/2$ response is symmetric and Lorentzian, 
whereas $\phi_0=0$ gives the antisymmetric dispersive response relevant to the simulations.
\textbf{b,} Analytical response of the ac driven junction from equation \ref{eq:mac2} for drive frequency $f=60\, \mHz$ and several scaled modulation strengths $\beta=u_1/(\hbar \omega)$, using $\Gamma/(2\pi) \approx 2\, \mHz$.
Increasing $\beta$ redistributes spectral weight among successive Shapiro harmonics and produces the characteristic dispersive resonance structure.
}
\label{Fig:EDmodel}
\end{figure*}

\textbf{Atomic junction response under dc and ac bias.} 
The dc potential-bias protocol involves a constant reservoir-energy offset $u_0$, 
applied suddenly at $t=0$ (Fig. \ref{Fig:EDdc}a). 
We characterize the simulation response through the imbalance  $\Delta N(t)=N_L(t)-N_R(t)$ and 
the junction phase difference $\phi_j(t)=\phi(x_0-3l,t)-\phi(x_0+3l,t)$,  
where $\phi(x,t)=\arg[\int dy\,\psi(x,y,t)/L_y]$ is obtained from the transversely averaged complex field. 
Fig. \ref{Fig:EDdc}b,c shows the evolution of $\Delta N(t)$ and $\phi_j(t)$ for 
biases $u_0/\mu =0$--$0.8$ over $166\, \mms$, where $\mu$ is the mean-field energy. 
Following an initial transient, the imbalance develops a pronounced response at small bias, 
whereas at intermediate and larger bias the net change becomes weaker and the junction enters a running-phase regime. 
In this regime, the wrapped phase $\phi_j(t)$ repeatedly traverses the $[-\pi, \pi]$ interval. 
As shown by the accompanying vortex dynamics (Fig. \ref{Fig:EDdc}d), 
these winding events are associated with vortex-mediated $2\pi$ phase slips.

 To obtain the dc current-bias characteristic, we calculate $I_\mdc = \langle  d(\Delta N)/dt \rangle_t$. 
For a barrier height $\tilde{V}_0=9$,  $I_{\rm dc}$ initially increases with $u_0$, reaches a maximum $I_m$ at a critical bias $u_c$, and subsequently decreases as the junction crosses into the running-phase regime (Fig. \ref{Fig:EDdc}e). 
The response therefore separates a low-bias, phase-locked Josephson regime from a high-bias dissipative running-phase regime.
These regimes are distinguished directly by the phase-winding frequency $\Omega_\phi=\langle\dot{\phi}_j\rangle_t$, extracted from the unwrapped phase.
Below $u_c$, $\Omega_\phi$ remains close to zero, indicating a phase-locked junction. 
Above $u_c$, $\Omega_\phi$ increases approximately linearly with the imposed bias, $\hbar\Omega_\phi\simeq u_0$, 
demonstrating the ac Josephson relation in the potential-biased geometry. 
The dc bias therefore directly controls the intrinsic phase-winding frequency of the running junction.

Because the initial junction phase difference is $\phi_j=0$, the Josephson current vanishes at $u_0=0$, and  
the finite-bias maximum in $I_\mdc$ develops dynamically from the subsequent phase evolution.
This dc-biased response is captured by the analytical model (equation \ref{eq:adc2}). 
Increasing the barrier height narrows the dc response, shifts its maximum to lower $u_0$, 
and reduces the peak current (Fig. \ref{Fig:EDdc}f). 
These trends reflect the simultaneous reduction of the critical current and 
of the effective damping scale governing the crossover into the running-phase regime.

A closely related broadening mechanism controls the ac-driven response. 
In that case, however, the transport is reorganized into dispersive Shapiro resonances near $u_0\simeq n\hbar\omega$, 
with $n=1,2,\ldots$ (Fig. \ref{Fig:EDac}a). 
Their characteristic width is likewise set by the dissipative dynamics of the junction: 
stronger damping at lower barrier heights broadens the resonances, 
whereas weaker damping at higher barriers sharpens them. 
Consistently, the total number of vortices nucleated during the driven dynamics 
decreases with increasing barrier height (Fig. \ref{Fig:EDac}c), 
supporting the connection between vortex-mediated phase relaxation and the effective damping of the junction. 
Because adjacent Shapiro resonances are separated in energy by $\hbar \omega$, 
their spectral resolution is controlled by the ratio of the resonance width to the drive frequency. 
At sufficiently strong damping, neighboring resonances broaden and begin to overlap, 
whereas reducing the damping resolves the individual Shapiro spikes more clearly.
These trends are captured by the analytical model (equation \ref{eq:mac2}), which links the finite resonance width and dispersive shape directly to the effective damping rate.

The resonance broadening depends not only on the barrier height but also on its width, 
as both parameters control the tunneling amplitude and hence the effective junction coupling. Their relative influence depends on the barrier geometries accessible in a given atomic platform. To illustrate this dependence, we show that spike resonances persist for a barrier height of $\tilde{V}_0=2$ at widths $w/\xi=4$ and $5$ (Fig. \ref{Fig:spikesVs2}). These results demonstrate the robustness of the resonance response to changes in barrier width and highlight the flexibility to optimize the trade-off between spectral resolution and signal strength across different atomic platforms \cite{SinghAJJ2025, Diego2026}.

\textbf{Collective Josephson junction model.} 
To capture the driven dynamics of the atomic Josephson junction, we treat the junction phase as a collective Josephson mode and incorporate dissipation through a phenomenological relaxation rate $\Gamma$.
We define the damping-weighted dc current as $I_\mdc = \Gamma \int_0^\infty e^{- \Gamma t} I_s(t) dt$, 
with $I_s(t) = I_c \sin \phi (t)$. 
For a static potential bias, the Josephson relation gives $\phi(t)=\phi_0+\omega_0 t$, 
where $\phi_0$ is the initial phase difference and $\omega_0=u_0/\hbar$.  
Evaluating the damping-weighted current yields
\begin{equation}\label{eq:adc}
\frac{I_\mdc}{I_c} = \frac{\Gamma^2 \sin \phi_0 + \Gamma \omega_0 \cos \phi_0}{\Gamma^2 + \omega_0^2}.
\end{equation}
The response therefore contains both symmetric Lorentzian and antisymmetric dispersive components (Fig. \ref{Fig:EDmodel}a). 
For $\phi_0=\pi/2$, the response is purely Lorentzian and becomes increasingly localized around zero bias as  $\Gamma$ decreases. For $\phi_0=0$, only the dispersive contribution remains, 
\begin{align}\label{eq:adc2}
\frac{I_\mdc}{I_c} =  \frac{\Gamma \omega_0}{\Gamma^2 + \omega_0^2}.
\end{align}
This form reproduces the characteristic finite-bias maximum followed by a decay in the simulated dc response (Fig.~\ref{Fig:EDdc}e,f). 
Equation~\ref{eq:adc2} reaches it maximum at $\omega_0= \Gamma$, with $I_\mdc=I_c/2$. 
The peak position therefore directly provides the characteristic relaxation scale $\hbar \Gamma  \simeq u_c$. 
We determine  $\Gamma$ by fitting equation \ref{eq:adc2} to the simulated response. 
Including a weak linear Ohmic contribution extends the agreement over the full bias range, while remaining negligible near the Josephson peak, showing that the peak response is dominated by the Josephson contribution.

Periodic modulation introduces an additional frequency scale in the Josephson dynamics, 
i.e., $\phi(t)=\phi_0+\omega_0 t - \beta\sin(\omega t)$, with $\beta=u_1/(\hbar\omega)$. 
Using the Jacobi-Anger expansion, the Josephson current can be written as 
\begin{align}
I_s = I_c \sum_{n=-\infty}^{\infty} J_n(\beta) \sin[\phi_0 + (\omega_0 - n \omega) t ],
\end{align}
which expands the phase modulation into harmonics of the drive. 
The corresponding damping-weighted dc current is  
\begin{equation}
\frac{I_\mdc}{I_c}  = \sum_{n=-\infty}^{\infty} J_n(\beta)  \frac{\Gamma^2 \sin \phi_0 + \Gamma (\omega_0- n \omega) \cos \phi_0}{\Gamma^2 + (\omega_0- n \omega)^2}. 
\end{equation}
For $\phi_0=0$, this reduces to 
\begin{equation}\label{eq:mac2}
\frac{I_\mdc}{I_c}  = \sum_{n} J_n(\beta)  \frac{\Gamma (\omega_0- n \omega)}{\Gamma^2 + (\omega_0- n \omega)^2}. 
\end{equation}
Equation~\ref{eq:mac2} directly captures the dispersive Shapiro response observed in the simulations (Fig.~\ref{Fig:EDac}a).
Each harmonic is centered at the resonance condition $\omega_0=n\omega$, 
where the corresponding contribution changes sign.
Its characteristic width is set by $\Gamma$, 
while its amplitude is controlled by $J_n(\beta)$.
For $|J_n(\beta)|$, the $n$-th contribution reaches a negative extremum at
$\omega_0=n\omega-\Gamma$,
vanishes at exact resonance, and reaches a positive extremum at
$\omega_0=n\omega+\Gamma$. 
The extrema have magnitude $I_c|J_n(\beta)|/2$.

The negative-current branch is particularly notable because it can occur while the imposed dc bias remains positive.
When the total simulated $I_{\rm dc}$ becomes negative, the junction therefore exhibits net particle transport opposite to the applied static bias, realizing a negative-mobility regime.
The role of $\Gamma$ is to give this resonant response a finite spectral width: 
weaker damping produces sharper, better-resolved Shapiro resonances, 
whereas stronger damping broadens neighboring resonances and can cause them to overlap.

To determine the effective damping rate, we fit Eq.~\ref{eq:mac2} to the simulated response for $\tilde V_0=9$ and $\beta=2.1$ (Fig.~\ref{Fig:EDac}b).
This yields $\Gamma/(2\pi)\simeq 2\,{\rm Hz}$.
Compared with the drive frequencies $f=60$--$90\,{\rm Hz}$, the relaxation rate is small, $\Gamma\ll\omega$, allowing individual Shapiro resonances to remain spectrally well resolved (Fig.~\ref{Fig:EDmodel}b).

\bibliography{References}

\end{document}